\documentclass[final,1p,times]{elsarticle} 
\usepackage{graphicx} 
\usepackage{amssymb} 
\usepackage{amsthm} 
\usepackage{lineno} 

\journal{Nuclear Physics A} 
\begin{document} 

\begin{frontmatter} 


\title{The QCD Phase Diagram from Chiral Approaches}

\author{Chihiro Sasaki}

\address{Physik-Department,
Technische Universit\"{a}t M\"{u}nchen,
D-85747 Garching, Germany}

\begin{abstract} 
I show an updated QCD phase diagram with recent developments
from chiral effective theories and phenomenological models.
Expected signals of a QCD critical point accessible in heavy-ion 
collisions are also discussed. In particular, non-monotonic 
behavior of fluctuations associated with conserved charges 
is focused on.
\end{abstract} 

\end{frontmatter} 


\section{Introduction}
\label{intro}

One of the main issues addressed in QCD is the phase structure
of strongly interacting matter at finite temperature
and baryon density.
Remarkable progress in lattice QCD simulations provides
a reliable description of bulk QCD matter for small
chemical potentials, i.e. equations of state and order of phase 
transitions, which is an input to heavy-ion 
phenomenologies~\cite{petreczky}.
Model studies of dense baryonic and quark matter have suggested 
a rich phase structure of QCD at temperatures and quark chemical 
potentials being of order $\Lambda_{\rm QCD}$.
Our knowledge on the QCD phase structure is however still limited:
The physics around normal nuclear matter density has been
empirically known and can be successfully described by
chiral effective field theories guided by experimental 
data~\cite{nuclearmatter}.
At asymptotically high density the Color-Flavor-Locked
phase is preferred as the QCD ground state~\cite{alford}.
In intermediate densities where lattice QCD is not accessible,
a description of dense matter still relies on effective
theories and models.

The order of the QCD phase transition is neither established
at low temperature or high density.
First-order phase transitions for cold dense matter
have been predicted in several approaches using 
chiral models~\cite{cp:models}, Schwinger-Dyson equations~\cite{cp:sd} 
and lattice QCD in strong coupling limit~\cite{cp:lat}.
This along with the observation of a crossover at zero chemical
potential from lattice QCD computations might suggest a critical
point in the QCD phase plane~\cite{cp:models}.
The existence of a QCD critical point is still an issue under 
debate. In fact, the location of a critical point strongly depends
on parameters, e.g. current quark masses and coupling strengths
of hadronic interactions in model studies~\cite{cp}. 

\begin{figure}
\begin{center}
\includegraphics[width=10cm]{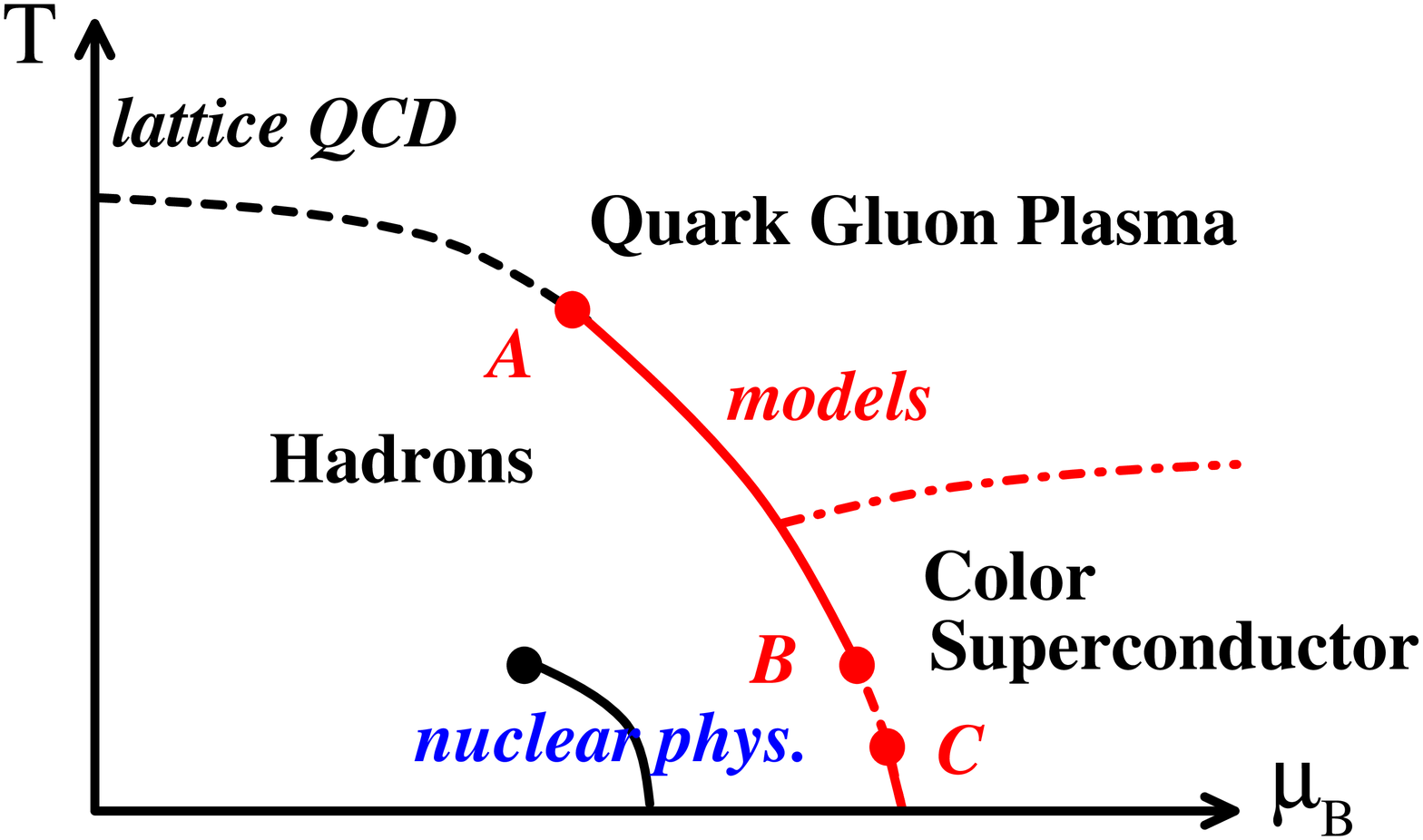}
\caption{
A sketch of the QCD phase diagram.
}
\label{phase}
\end{center}
\end{figure}

The presence of diquarks in dense matter leads to a possibility
of new critical points at low temperature depending on 
couplings of the chiral condensate to diquarks~\cite{cps1,cps2}.
A hypothetical phase diagram is shown in Figure~\ref{phase}
assuming multiple critical points A-C.
Since the critical point C appears in the two-flavored
Nambu--Jona-Lasinio (NJL) model with a fine-tuning of the relative 
strength of the chiral and diquark couplings~\cite{cps2},
this might disappear from the phase diagram due to correlations
via pion exchanges in dense medium.
The appearance of the point B indicates a crossover from hadronic matter
to color superconductor. This is a realization of hadron-quark 
continuity~\cite{cont} in the Ginzburg-Landau description.
Close similarities to the BEC-BCS crossover in ultra-cold atomic
systems are also suggestive~\cite{baym}.

\section{Chiral symmetry breaking vs. confinement}
\label{conf}

\begin{figure}
\begin{center}
\includegraphics[width=11cm]{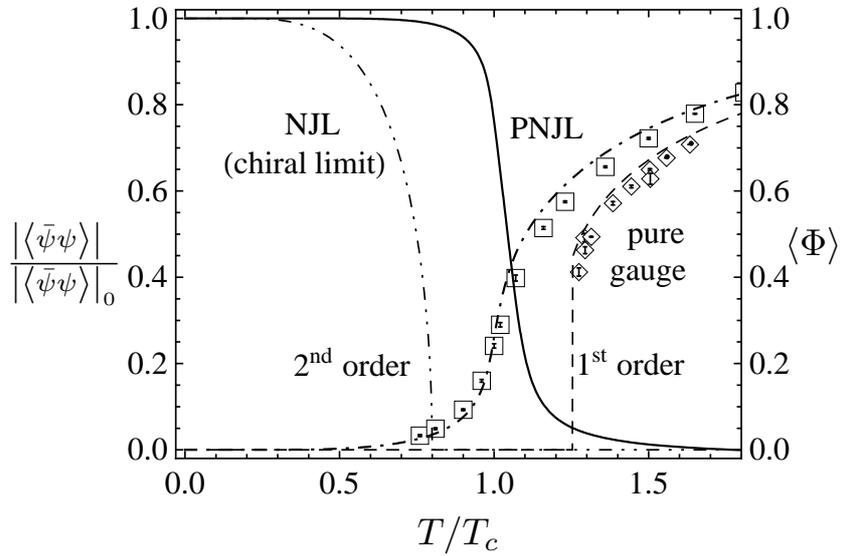}
\caption{
The chiral condensate normalized to its vacuum value
(solid line) and the Polyakov loop (dashed dotted line)
in a PNJL model~\cite{pnjl}. The data from
lattice calculations  for the Polyakov loop in pure gauge 
and in full QCD are shown in~\cite{lattice}.
}
\label{pnjl}
\end{center}
\end{figure}

Strong interaction leads not only to dynamical breaking of
chiral symmetry but also to confinement.
Both features are characterized by strict order parameters
associated with global symmetries of the QCD Lagrangian
in two limiting situations:
the quark bilinear $\langle \bar{q}q \rangle$ in the limit
of massless quarks, and the Polyakov loop $\langle \Phi \rangle$ 
in the limit of infinitely heavy quarks.
The NJL model with Polyakov loops (PNJL model) has been developed
to deal with those features simultaneously~\cite{pnjl}.
The model describes that only three-quark states are thermally
relevant below the chiral critical temperature, which is
reminiscent of confinement.
The two order parameters in the PNJL model, chiral condensate
and Polyakov loop expectation values, are shown in Figure~\ref{pnjl}.
Due to finite quark masses chiral and deconfinement transitions
are crossover and the corresponding pseudo-critical temperatures,
$T_{\rm ch}$ and $T_{\rm dec}$, are defined as the steepest points 
of derivatives of $\langle \bar{q}q \rangle$ and $\langle \Phi \rangle$.
The interference of quark with gluon sectors makes
both $T_{\rm ch}$ and $T_{\rm dec}$ mutually shifted.
At zero chemical potential one sees that the two transitions 
take place almost simultaneously, as indicated in lattice QCD.

\begin{figure}
\begin{center}
\includegraphics[width=12cm]{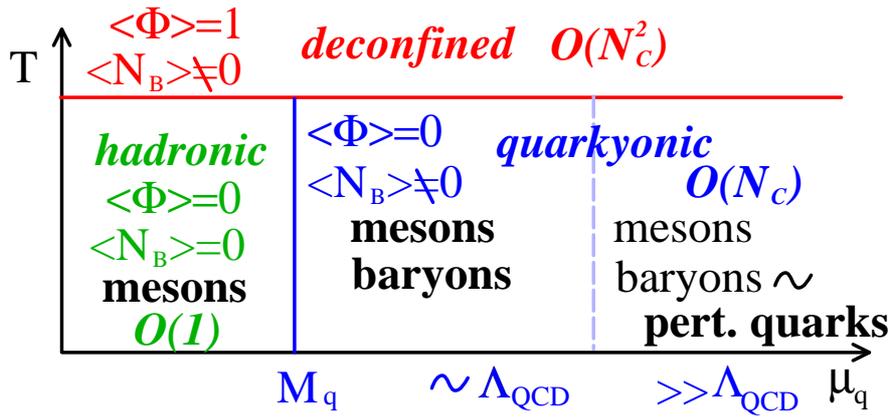}
\caption{
The phase diagram in large $N_c$ proposed in~\cite{quarkyonic}.
}
\label{largenc}
\end{center}
\end{figure}

The phase transitions at higher chemical potential may have
a close relation to the notion of Quarkyonic Phase which
has been proposed as a novel phase of dense quarks based on the 
argument using large $N_c$ counting where $N_c$ denotes number of 
colors~\cite{quarkyonic}:
In the large $N_c$ limit there are three phases which are
rigorously distinguished using $\langle \Phi \rangle$ and the 
baryon number density $\langle N_B \rangle$. The quarkyonic phase 
is characterized by $\langle \Phi \rangle = 0$ indicating the 
system confined and non-vanishing $\langle N_B \rangle$ above 
$\mu_B = M_B$ with a baryon mass $M_B$. The phase structure
in large $N_c$ is shown in Figure~\ref{largenc}.
A possible deformation of the phase boundaries in Figure~\ref{largenc} 
together with chiral phase transition is illustrated using a PNJL 
model~\cite{mrs}. 
\begin{figure}
\begin{center}
\includegraphics[width=11cm]{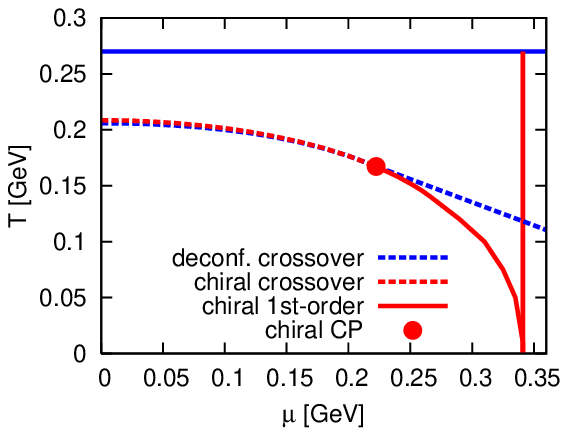}
\caption{
The phase diagram of a PNJL model for different $N_c$~\cite{mrs}. 
Two straight lines indicate the deconfinement and chiral phase 
transitions for $N_c = \infty$ and the lower curves for $N_c=3$.
}
\label{phasepnjl}
\end{center}
\end{figure}
Finite $N_c$ corrections make the transition lines bending down. 
One finds that for $N_c=3$ deconfinement and chiral crossover lines 
are on top of each other in a wide range of $\mu$. A critical point 
associated with chiral symmetry appears around the junction of those 
crossovers.

The separation of the quarkyonic from hadronic phase is not clear 
any more in a system with finite $N_c$. Nevertheless, a rapid change 
in the baryon number would be interpreted as the quarkyonic transition 
which separates meson dominant from baryon dominant regions. This 
might appear near the boundary for chemical equilibrium where one 
would expect a rapid change in the number of degrees of 
freedom~\cite{quarkyonic}.
More realistic models including mesons and baryons, rather than chiral 
quarks, are indispensable to further understanding of the physics 
of dense baryonic matter and possible appearance of the quarkyonic 
``phase'' in $N_c=3$ QCD. 
In particular, chiral symmetry restoration for baryons must
be worked out. Although two alternatives for chirality assignment 
to baryons have been known, it remains an open question which
scenario is preferred by nature~\cite{pdoubling}.

\section{Fluctuations and critical points}
\label{fluc}

Modifications in the magnitude of fluctuations or the corresponding 
susceptibilities are considered as a possible signal for
deconfinement and chiral symmetry restoration.
In this context, fluctuations related to conserved charges play an 
important role since they are directly accessible in 
experiments~\cite{fluct}.
Especially, non-monotonic behavior of baryon number fluctuations 
could be a clear indication for the existence of a critical point 
in the QCD phase diagram.
Higher moments and their signs have also been proposed as more
sensitive probes to the phase transitions~\cite{masky}.
If the critical region is sufficiently large, 
the critical point could be an attractor
of isentropes and lead to a change in the transverse velocity 
dependence of proton-antiproton ratio~\cite{nonaka}.
However, a model study has recently suggested that such strong 
modifications are washed out by quantum fluctuations~\cite{eiji} 
and this indicates no focusing.
Besides, isentropes are not universal since entropy and
baryon number densities have no derivatives of the order parameter
which are responsible for the universality.

The suppression of density fluctuations along the 
first-order transition appears under the assumption that this 
transition takes place in equilibrium. This is modified when 
there is a deviation from equilibrium~\cite{our:spinodal}.
When entering the coexistence region, a singularity in the
net quark number susceptibility $\chi_q$ appears at the isothermal
spinodal lines, where the fluctuations diverge and the 
susceptibility changes sign. 
In between the spinodal lines, the susceptibility is negative. 
This implies  instabilities in the baryon number 
fluctuations when crossing from a meta-stable to an unstable phase.
The above behavior of $\chi_q$ is a direct consequence of the 
thermodynamics relation,
$
\left( \frac{\partial P}{\partial V} \right)_T
= - \frac{n_q^2}{V}\frac{1}{\chi_q}
$\,.
Along the isothermal spinodals the pressure derivative vanishes.
Thus, for non-vanishing density $n_q$, $\chi_q$ must diverge to 
satisfy this relation.
\begin{figure}
\begin{center}
\includegraphics[width=11cm]{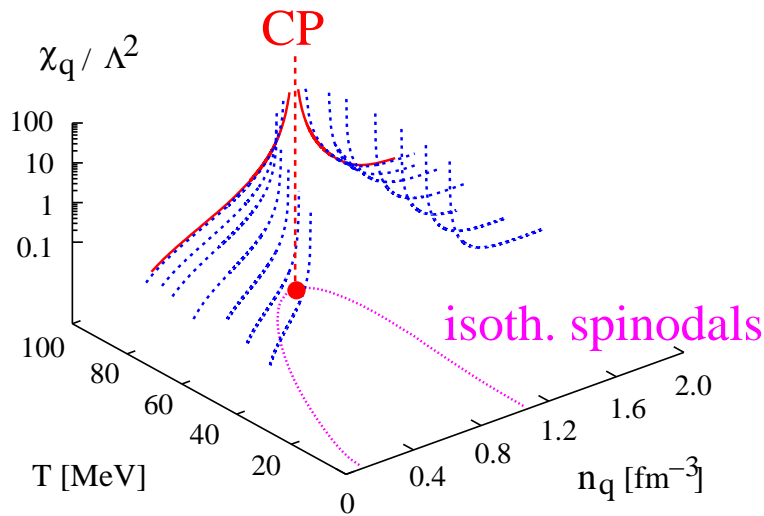}
\caption{
The net quark number susceptibility $\chi_q$ in the stable
and meta-stable regions~\cite{our:spinodal}.
Here $\chi_q$ is normalized to the three-momentum cutoff 
$\Lambda = 587.9$ MeV used in the $N_f=2$ NJL model.
}
\label{chiq}
\end{center}
\end{figure}
The evolution of the singularity at the spinodal lines in the $T$-$n_q$ 
plane under the mean field approximation
is shown in Figure~\ref{chiq}. The critical 
exponent at the isothermal spinodal line is found to be $\gamma=1/2$, 
with $\chi_q \sim (\mu-\mu_c)^{-\gamma}$, while $\gamma=2/3$ at the 
critical point~\cite{our:spinodal}. Thus, the singularities at the 
two spinodal lines conspire to yield a somewhat stronger divergence 
as they join at the critical point. The critical region of enhanced 
susceptibility around the critical point is found to be fairly small 
in and beyond the mean field approximation~\cite{schaefer-wambach}, 
while in the more realistic non-equilibrium system one expects 
fluctuations in a larger region of the phase diagram, i.e.
over a broader range of beam energies, due to the spinodal instabilities.

\section{Conclusions}
\label{conclusions}

Our understanding of the QCD phases remains inadequate
to make a firm statement about the phase diagram and
the existence of one or more critical points. 
Model studies carried out under the mean field approximation
are expected to capture some essential features associated with
chiral dynamics. However, since under this approximation one omits 
important effects, e.g. quantum fluctuations, from a theory with 
a limited 
number of degrees of freedom, the models could lead to even 
qualitatively a different result from QCD.

\begin{figure}
\begin{center}
\includegraphics[width=10cm]{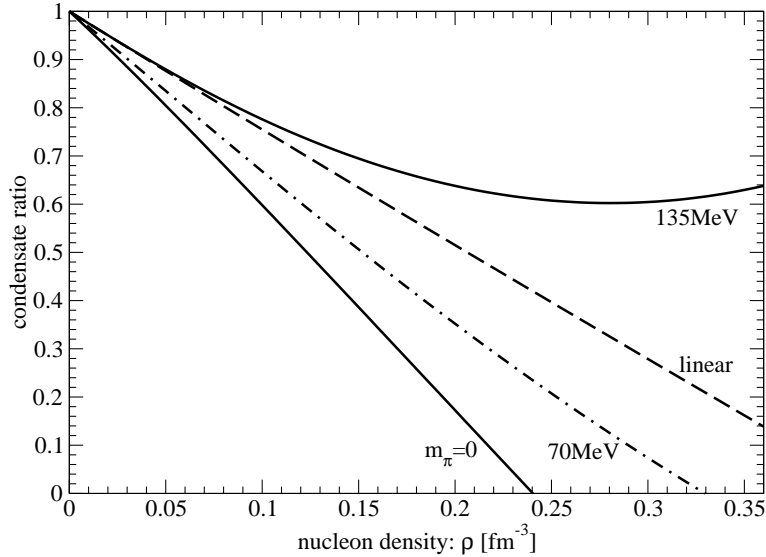}
\caption{
The ratio of the in-medium chiral condensate to its
vacuum value as a function of the nucleon density
for several pion masses, $m_\pi = 0\,,70\,,135$ MeV~\cite{chpt}.
}
\label{chcond}
\end{center}
\end{figure}

Some lessons for more realistic modeling can be found in chiral 
effective field theories (EFTs) applied to normal nuclear matter: 
The nuclear equation of state (EoS) for various nuclei and the 
liquid-gas phase transition have been explored within the chiral 
approaches which allow us to make quantitative 
science~\cite{nuclearmatter}.
Figure~\ref{chcond} shows the in-medium chiral condensate
normalized to its vacuum value calculated in the chiral EFT
constrained by nuclear EoS~\cite{chpt}.
One observes that the in-medium condensate strongly relies on
the pion mass. This comes from two-pion exchange correlations
with virtual $\Delta(1232)$ excitations and stabilizes
the dropping of the condensate for physical pion mass with 
increasing density. This indicates that the chiral symmetry
restoration is delayed and would take place much higher density
than $2\,\rho_0$. This also could suggest that in-medium 
correlations make a phase transition obscure and eventually
first-order phase transition might disappear from the phase diagram.
On the other hand, in the chiral limit 
the condensate vanishes already at $\sim 1.5\,\rho_0$.
It is remarkable that small quark masses, $m_{u,d} \sim 5$ MeV, 
influence over thermodynamics so strongly.
Obviously, a more realistic description for hadronic matter at higher 
densities and the critical point(s) should be provided 
in systematic approaches with mesons and baryons.

In condensed matter physics strongly coupled systems can be
described by EFTs where the notion of
e.g. quasi-particles and BCS pairings has been successful.
One would thus expect a similarity in nuclear many-body systems 
in high densities. In fact, the spectroscopic factor for various nuclei
indicating a deviation from the fully occupied mean-field 
orbits clearly shows that single-particle-ness of excitations
reaches $\sim$70 \%, see e.g. Figure 11 in~\cite{spectroscopic}.
This would encourage modeling dense baryonic matter
in a quasi-particle picture.

\section*{Acknowledgments}

I am grateful for fruitful discussions with L.~McLerran,
K.~Redlich, M.~Rho and W.~Weise. 
The work has been supported in part by the DFG cluster 
of excellence ``Origin and Structure of the Universe''.




\begin{thebibliography}{00} 
   
\bibitem{petreczky}
P.~Petreczky, these proceedings.

\bibitem{nuclearmatter}
see e.g., 
P.~Finelli, N.~Kaiser, D.~Vretenar and W.~Weise,
  Nucl.\ Phys.\  A {\bf 735}, 449 (2004);
  Nucl.\ Phys.\  A {\bf 770}, 1 (2006),
S.~Fritsch, N.~Kaiser and W.~Weise,
  Nucl.\ Phys.\  A {\bf 750}, 259 (2005).

\bibitem{alford}
M.~Alford, these proceedings.

\bibitem{cp:models}
  M.~Asakawa and K.~Yazaki,
  Nucl.\ Phys.\  A {\bf 504}, 668 (1989),
  J.~Berges and K.~Rajagopal,
  Nucl.\ Phys.\  B {\bf 538}, 215 (1999),
  A.~M.~Halasz, A.~D.~Jackson, R.~E.~Shrock, M.~A.~Stephanov 
  and J.~J.~M.~Verbaarschot,
  Phys.\ Rev.\  D {\bf 58}, 096007 (1998),
  O.~Scavenius, A.~Mocsy, I.~N.~Mishustin and D.~H.~Rischke,
  Phys.\ Rev.\  C {\bf 64}, 045202 (2001),
  H.~Fujii,
  Phys.\ Rev.\ D {\bf 67}, 094018 (2003),
  H.~Fujii and M.~Ohtani,
  Phys.\ Rev.\ D {\bf 70}, 014016 (2004).

\bibitem{cp:sd}
A.~Barducci, R.~Casalbuoni, S.~De Curtis, R.~Gatto and G.~Pettini,
  Phys.\ Lett.\  B {\bf 231}, 463 (1989);
  Phys.\ Rev.\  D {\bf 41}, 1610 (1990),
Y.~Taniguchi and Y.~Yoshida,
  Phys.\ Rev.\  D {\bf 55}, 2283 (1997),
M.~Harada and A.~Shibata,
  Phys.\ Rev.\  D {\bf 59}, 014010 (1999).

\bibitem{cp:lat}
E.~M.~Ilgenfritz and J.~Kripfganz,
  Z.\ Phys.\  C {\bf 29}, 79 (1985),
P.~H.~Damgaard, D.~Hochberg and N.~Kawamoto,
  Phys.\ Lett.\  B {\bf 158}, 239 (1985),
F.~Karsch and K.~H.~Mutter,
  Nucl.\ Phys.\  B {\bf 313}, 541 (1989),
N.~Kawamoto, K.~Miura, A.~Ohnishi and T.~Ohnuma,
  Phys.\ Rev.\  D {\bf 75}, 014502 (2007).

\bibitem{cp}
S.~Klimt, M.~Lutz and W.~Weise,
  Phys.\ Lett.\  B {\bf 249}, 386 (1990),
Y.~Hatta and T.~Ikeda,
  Phys.\ Rev.\  D {\bf 67}, 014028 (2003),
 B.~J.~Schaefer and J.~Wambach,
  Nucl.\ Phys.\  A {\bf 757}, 479 (2005),
  P.~de Forcrand and O.~Philipsen,
  JHEP {\bf 0701}, 077 (2007).

\bibitem{cps1}
M.~Kitazawa, T.~Koide, T.~Kunihiro and Y.~Nemoto,
  Prog.\ Theor.\ Phys.\  {\bf 108}, 929 (2002),
 T.~Hatsuda, M.~Tachibana, N.~Yamamoto and G.~Baym,
  Phys.\ Rev.\ Lett.\  {\bf 97}, 122001 (2006),
N.~Yamamoto, M.~Tachibana, T.~Hatsuda and G.~Baym,
  Phys.\ Rev.\  D {\bf 76}, 074001 (2007).

\bibitem{cps2}
Z.~Zhang, K.~Fukushima and T.~Kunihiro,
  Phys.\ Rev.\  D {\bf 79}, 014004 (2009).

\bibitem{cont}
  T.~Schafer and F.~Wilczek,
  Phys.\ Rev.\ Lett.\  {\bf 82}, 3956 (1999).

\bibitem{baym}
  G.~Baym, T.~Hatsuda, M.~Tachibana and N.~Yamamoto,
  J.\ Phys.\ G {\bf 35}, 104021 (2008).

\bibitem{pnjl}
K.~Fukushima,
  Phys.\ Lett.\  B {\bf 591}, 277 (2004),
C.~Ratti, M.~A.~Thaler and W.~Weise,
  Phys.\ Rev.\  D {\bf 73}, 014019 (2006),
E.~Megias, E.~Ruiz Arriola and L.~L.~Salcedo,
  Phys.\ Rev.\  D {\bf 74}, 065005 (2006),
S.~K.~Ghosh, T.~K.~Mukherjee, M.~G.~Mustafa and R.~Ray,
  Phys.\ Rev.\  D {\bf 73}, 114007 (2006),
S.~Roessner, C.~Ratti and W.~Weise,
  Phys.\ Rev.\  D {\bf 75}, 034007 (2007),
C.~Ratti, S.~Roessner and W.~Weise,
  Phys.\ Lett.\  B {\bf 649}, 57 (2007),
C.~Sasaki, B.~Friman and K.~Redlich,
  Phys.\ Rev.\  D {\bf 75}, 074013 (2007),
H.~Hansen, W.~M.~Alberico, A.~Beraudo, A.~Molinari, M.~Nardi 
and C.~Ratti,
  Phys.\ Rev.\  D {\bf 75}, 065004 (2007),
Z.~Zhang and Y.~X.~Liu,
  Phys.\ Rev.\  C {\bf 75}, 064910 (2007),
S.~Roessner, T.~Hell, C.~Ratti and W.~Weise,
  Nucl.\ Phys.\  A {\bf 814}, 118 (2008),
K.~Kashiwa, H.~Kouno, M.~Matsuzaki and M.~Yahiro,
  Phys.\ Lett.\  B {\bf 662}, 26 (2008),
Y.~Sakai, K.~Kashiwa, H.~Kouno and M.~Yahiro,
  Phys.\ Rev.\  D {\bf 77}, 051901 (2008);
  Phys.\ Rev.\  D {\bf 78}, 036001 (2008),
H.~Abuki, M.~Ciminale, R.~Gatto, G.~Nardulli and M.~Ruggieri,
  Phys.\ Rev.\  D {\bf 77}, 074018 (2008);
  Phys.\ Rev.\  D {\bf 78}, 034034 (2008),
K.~Fukushima,
  Phys.\ Rev.\  D {\bf 77}, 114028 (2008)
  [Erratum-ibid.\  D {\bf 78}, 039902 (2008)];
  arXiv:0901.0783 [hep-ph],
  H.~Abuki, M.~Ciminale, R.~Gatto and M.~Ruggieri,
  Phys.\ Rev.\  D {\bf 79}, 034021 (2009),
T.~Hell, S.~Roessner, M.~Cristoforetti and W.~Weise,
  Phys.\ Rev.\  D {\bf 79}, 014022 (2009),
H.~Abuki and K.~Fukushima,
  Phys.\ Lett.\  B {\bf 676}, 57 (2009).

\bibitem{lattice}
  O.~Kaczmarek and F.~Zantow,
  Phys.\ Rev.\  D {\bf 71}, 114510 (2005).

\bibitem{quarkyonic}
  L.~McLerran and R.~D.~Pisarski,
  Nucl.\ Phys.\  A {\bf 796}, 83 (2007),
  Y.~Hidaka, L.~D.~McLerran and R.~D.~Pisarski,
  Nucl.\ Phys.\  A {\bf 808}, 117 (2008).

\bibitem{mrs}
  L.~McLerran, K.~Redlich and C.~Sasaki,
  Nucl.\ Phys.\  A {\bf 824}, 86 (2009).

\bibitem{pdoubling}
  C.~E.~Detar and T.~Kunihiro,
  Phys.\ Rev.\  D {\bf 39}, 2805 (1989),
Y.~Nemoto, D.~Jido, M.~Oka and A.~Hosaka,
  Phys.\ Rev.\  D {\bf 57}, 4124 (1998),
D.~Jido, Y.~Nemoto, M.~Oka and A.~Hosaka,
  Nucl.\ Phys.\  A {\bf 671}, 471 (2000),
H.~c.~Kim, D.~Jido and M.~Oka,
  Nucl.\ Phys.\  A {\bf 640}, 77 (1998),
D.~Jido, T.~Hatsuda and T.~Kunihiro,
  Phys.\ Rev.\ Lett.\  {\bf 84}, 3252 (2000).

\bibitem{fluct}
  M.~A.~Stephanov, K.~Rajagopal and E.~V.~Shuryak,
  Phys.\ Rev.\ Lett.\  {\bf 81}, 4816 (1998),
 S. Jeon and V. Koch,
{\it Quark Gluon Plasma 3}, Eds. R.C. Hwa and
X. N. Wang, World Scientific Publishing, 2004.

\bibitem{masky}
  S.~Ejiri, F.~Karsch and K.~Redlich,
  Phys.\ Lett.\  B {\bf 633}, 275 (2006),
B.~Stokic, B.~Friman and K.~Redlich,
  Phys.\ Lett.\  B {\bf 673}, 192 (2009),
M.~A.~Stephanov,
  Phys.\ Rev.\ Lett.\  {\bf 102}, 032301 (2009),
  M.~Asakawa, S.~Ejiri and M.~Kitazawa,
  arXiv:0904.2089 [nucl-th].

\bibitem{nonaka}
  C.~Nonaka and M.~Asakawa,
  Phys.\ Rev.\  C {\bf 71}, 044904 (2005),
M.~Asakawa, S.~A.~Bass, B.~Muller and C.~Nonaka,
  Phys.\ Rev.\ Lett.\  {\bf 101}, 122302 (2008).

\bibitem{eiji}
  E.~Nakano, B.~J.~Schaefer, B.~Stokic, B.~Friman and K.~Redlich,
  arXiv:0907.1344 [hep-ph].

\bibitem{our:spinodal}
  C.~Sasaki, B.~Friman and K.~Redlich,
  Phys.\ Rev.\ Lett.\  {\bf 99}, 232301 (2007);
  Phys.\ Rev.\  D {\bf 77}, 034024 (2008).

\bibitem{schaefer-wambach}
  B.~J.~Schaefer and J.~Wambach,
  Phys.\ Rev.\  D {\bf 75}, 085015 (2007);
C.~Sasaki, B.~Friman and K.~Redlich,
  Phys.\ Rev.\  D {\bf 75}, 054026 (2007).

\bibitem{chpt}
  N.~Kaiser, P.~de Homont and W.~Weise,
  Phys.\ Rev.\  C {\bf 77}, 025204 (2008).

\bibitem{spectroscopic}
W.~H.~Dickhoff and C.~Barbieri,
  Prog.\ Part.\ Nucl.\ Phys.\  {\bf 52}, 377 (2004).

\end{thebibliography}
\end{document}